\documentclass[%
 reprint,
superscriptaddress,
 amsmath,
 amssymb,
 aps, 
floatfix,
]{revtex4-2}

\usepackage{graphicx}
\usepackage{dcolumn}
\usepackage{bm}
\usepackage{xcolor} 

\usepackage[colorlinks=true, allcolors=blue, breaklinks=true]{hyperref}

\begin{document}

\title{\textbf{Superconducting Qubit Readout Using Next-Generation Reservoir Computing} 
}%

\author{Robert Kent}
\affiliation{The Ohio State University, Department of Physics, 191 West Woodruff Ave., Columbus, OH 43210, USA.}

\author{Benjamin Lienhard}
\affiliation{Department of Chemistry, Princeton University, Princeton, NJ 08544, USA}
\affiliation{Department of Electrical and Computer Engineering, Princeton University, Princeton, NJ 08544, USA}

\author{Gregory Lafyatis}
\affiliation{The Ohio State University, Department of Physics, 191 West Woodruff Ave., Columbus, OH 43210, USA.}

\author{Daniel J. Gauthier}
\email{Contact author: daniel.gauthier62@gmail.com}
\affiliation{ResCon Technologies, LLC, 1275 Kinnear Rd., Suite 239, Columbus, OH 43212, USA}

\noaffiliation

\date{\today}

\begin{abstract}
Quantum processors require rapid and high-fidelity simultaneous measurements of many qubits. While superconducting qubits are among the leading modalities toward a useful quantum processor, their readout remains a bottleneck. Traditional approaches to processing measurement data often struggle to account for crosstalk present in frequency-multiplexed readout, the preferred method to reduce the resource overhead. Recent approaches to address this challenge use neural networks to improve the state-discrimination fidelity. However, they are computationally expensive to train and evaluate, resulting in increased latency and poor scalability as the number of qubits increases. We present an alternative machine learning approach based on next-generation reservoir computing that constructs polynomial features from the measurement signals and maps them to the corresponding qubit states. This method is highly parallelizable, avoids the costly nonlinear activation functions common in neural networks, and supports real‑time training, enabling fast evaluation, adaptability, and scalability. Despite its lower computational complexity, our reservoir approach is able to maintain high qubit-state-discrimination fidelity. Relative to traditional methods, our approach achieves error reductions of up to 50\% and 11\% on single- and five-qubit datasets, respectively, and delivers up to 2.5× crosstalk reduction on the five-qubit dataset. Compared with recent machine-learning methods, evaluating our model requires 100× fewer multiplications for single-qubit and 2.5× fewer for five-qubit models. This work demonstrates that reservoir computing can enhance qubit-state discrimination while maintaining scalability for future quantum processors.
\end{abstract}

\maketitle


\section{\label{sec:Introduction} Introduction}

Quantum computers have the potential to solve certain problems exponentially faster than classical computers---a capability that could be used to break encryption algorithms~\cite{Shor_1994,Bernstein_2017,Gidney_2025}, discover new drugs~\cite{Cao_2018,Blunt_2022}, or perform complex optimization tasks~\cite{Farhi_2014,Zhou_2020,optimization_nature_2024}. However, realizing this computational advantage will require scaling beyond today’s small-scale quantum processors comprising tens to a few hundred qubits and reducing the error rates.  Although quantum error correction algorithms offer a solution to detect and correct errors~\cite{Shor_1995,Fowler_2012,Google_2024}, they are based on rapid and high-fidelity mid-circuit measurements. Thus, it is essential to develop fast, accurate, and scalable qubit readout strategies.


A leading quantum computing architecture is based on superconducting qubits. Here, the state of the qubit ($|0\rangle$ or $|1 \rangle$) is measured in the dispersive regime, in which a resonator dedicated to measuring the qubit acquires a frequency shift $\chi$ of up to a few megahertz that depends on the state of the qubit. This shift is measured by probing the resonator with a coherent microwave pulse. The reflected or transmitted signal comprises the qubit-state information imprinted in the phase and amplitude. The signal in the range of a few gigahertz is frequency downconverted to the megahertz regime by mixing it with a local oscillator to extract the in-phase $(\mathcal{I}_n)$ and `quadrature' $(\mathcal{Q}_n)$ components at time step $n$ during a measurement window. These signals are digitized by control hardware, such as a field-programmable gate array (FPGA), and processed with an algorithm to infer the qubit's state.

\begin{figure*}[bt]
\includegraphics[width=\textwidth]{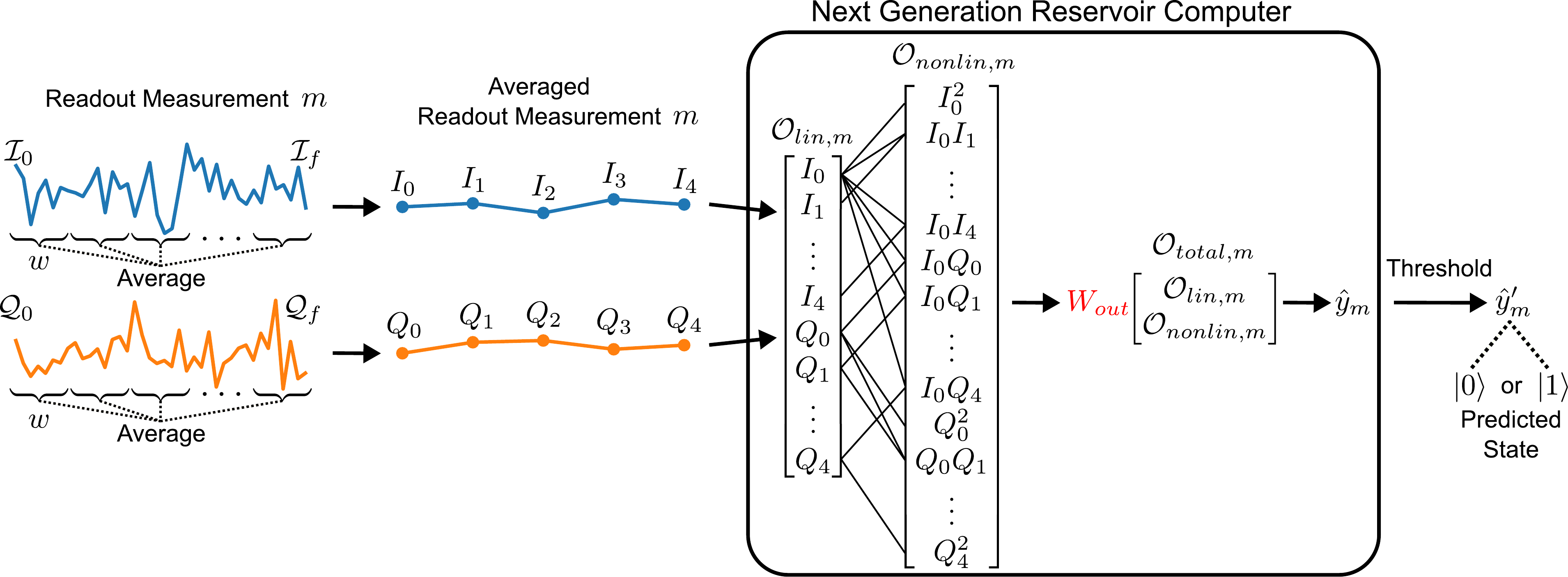}
\caption{\label{fig:storyboard} Superconducting qubit readout using the next-generation reservoir computer (NG-RC) algorithm followed by a discriminator.}
\end{figure*}

Designing qubit-state readout algorithms with high classification fidelity remains a challenge due to errors such as an excited state relaxing to the ground state, excitations to a higher state, and experimental noise, limiting signal-to-noise ratio (SNR). Additionally, qubits are often measured simultaneously using a frequency-multiplexed data bus, which can suffer from crosstalk between the individual qubit-readout signals, resonators, and qubits.

Traditional readout-signal filters, such as boxcar filters or matched filters (see details in Sec.~\ref{sec: Traditional approaches to qubit discrimination}), are computationally efficient, making them scalable for larger qubit systems. However, they lack nonlinear processing, which can increase accuracy. Furthermore, as they are often restricted from discriminating more than two states, their ability to account for crosstalk or higher order states is limited. 

Specific machine-learning (ML) tools can address some of these readout-signal processing challenges. One recent study improves upon matched filtering by using linear regression to optimize filter weights~\cite{Khan_2024}, but it has not been tested on multi-qubit systems. Other studies use feed-forward neural networks in conjunction with different signal processing routines~\cite{Lienhard_2022, Satvik_2023, Gautam_2024} to account for crosstalk and improve the qubit-state-discrimination fidelity, but are computationally expensive to train. Furthermore, their complexity increases latency and hardware demands, limiting their scalability to larger systems.

Here, we introduce a new ML readout algorithm based on next-generation reservoir computing (NG-RC)~\cite{Gauthier_2021}, illustrated in Fig.~\ref{fig:storyboard}, it combines the computational efficiency and scalability of traditional approaches with the high classification fidelity typical of large neural network models.

Briefly, the $(\mathcal{I}_n,\mathcal{Q}_n)$ data for a single measurement $m$ are averaged over non-overlapping time windows of size $w$ to form linear $O_{lin,m}$ and nonlinear functions $O_{nonlin,m}$ of the input variables. These variables are the components of the so-called \textit{feature vector}. Here, we consider low-order polynomial functions for the nonlinear features because they encode low-order correlations in the data. The features are multiplied by weights collected in a matrix $W_{out}$, summed to produce the output $\hat{y}_m$, and a threshold is applied to obtain the predicted qubit state $\hat{y}'_m \in \{0,1\}$. The model is \textit{trained} using supervised learning, where $y_m$ is set to the known initialized qubit state, and $W_{out}$ is found through regularized linear regression. Once the NG-RC model is trained, it can be used to predict the qubit state on data it has not previously encountered.

We evaluated our approach using single-qubit datasets used by Khan~\textit{et al.}~\cite{Khan_2024} and the five-qubit dataset analyzed by Lienhard~\textit{et al.}~\cite{Lienhard_2022,Satvik_2023}. We compare the qubit-state-discrimination fidelity of the NG-RC algorithm to their approaches as well as to traditional (mode) matched filters.


\section{\label{sec:Background} Background}

\subsection{ \label{sec: Traditional approaches to qubit discrimination} Traditional Approaches to Qubit-State Discrimination}

The boxcar filter and subsequent thresholding is one of the simplest and most widely used methods for qubit-state discrimination. Here, the readout trace bins are summed over a time window of interest. The filtered signal is represented by a complex number in the $\mathcal{I}\mathcal{Q}$-plane as
\begin{equation}
\label{eq:boxcar}
    \mathcal{S}_{\text{boxcar}} = \sum_{n}^N [S]_n = \sum_{n}^N (\mathcal{I}_n + i \mathcal{Q}_n),
\end{equation}
$i=\sqrt{-1}$ is the imaginary unit. A discriminator is applied to the filtered value to infer the qubit state. 

The matched filter extends the boxcar filter by introducing complex weights $k_n$ as
\begin{equation}
\label{eq:matched filter}
    \mathcal{S}_{\text{MF}} = \sum_{n}^N k_n[S]_n = \sum_{n}^N k_n (\mathcal{I}_n + i \mathcal{Q}_n),
\end{equation}
which are chosen to maximize the SNR, which is given by 
\begin{equation}
\label{eq:matched filter weights}
    k_n = \frac{\langle [S_0]_n - [S_1]_n\rangle}{\text{var}([S_0]_n) + \text{var}([S_1]_n)}
\end{equation}
for two-class discrimination. Here, $S_0$ and $S_1$ are the signals for the ground and excited states, respectively, $\langle ... \rangle$ denotes the mean value, and $\text{var}(...)$ the variance. Eq.~\ref{eq:matched filter weights} assumes Gaussian-distributed noise in the signal, which is not always accurate~\cite{Khan_2024}. Like the boxcar filter, a discriminator is applied to the filtered value to infer the qubit state. 

The boxcar and matched filter typically use the individual qubit signals, extracted through digital demodulation of the frequency-multiplexed signal at the corresponding intermediate frequency. While this results in an additional computational step, these approaches typically demand substantially fewer operations than ML-based algorithms.

\subsection{\label{sec: Machine learning approaches to qubit discrimination}  Machine-Learning Approaches to Qubit-State Discrimination}




Recently, Khan~\textit{et al.}~\cite{Khan_2024} demonstrated that a simple ML-based algorithm, which they call the trainable temporal postprocessor (TPP), can outperform traditional methods by avoiding assumptions about noise distribution and classifying an arbitrary number of states. Their approach determines a linear combination of $\mathcal{I}_n$ and $\mathcal{Q}_n$ at all times $t_n$ in the readout signal that maps to the predicted qubit state. The output is a one-hot encoding, such as $[1,0]$ for $|0\rangle$ and $[0,1]$ for $|1\rangle$, where the index with the higher value represents the predicted qubit state. The trained TPP algorithm uses weights obtained via linear regression, one of the simplest supervised learning approaches. It has a complexity similar to a matched filter, making it efficient to evaluate during deployment. However, this model lacks nonlinear processing, which can limit its classification performance and its ability to mitigate crosstalk in frequency-multiplexed qubit-state readout. 

In contrast, other recent ML-based algorithms leverage feedforward neural networks (FNNs)~\cite{Lienhard_2022,Satvik_2023,Gautam_2024,Reuer_2023}. An FNN consists of an input layer, followed by a series of `hidden layers' comprised of nodes or `neurons' that perform a weighted sum followed by a nonlinear activation function. The output layer encodes the qubit labels. 

These FNNs are trained using supervised learning, which involves nonlinear optimization algorithms because the weights are embedded within nonlinear functions. This process is computationally demanding, often requiring hours of processing time on graphical processing units (GPUs). Additionally, these models frequently have millions of parameters, increasing the computational optimization cost. As a result, they cannot quickly adapt to changing experimental conditions, such as those induced between cooldown cycles of the dilution refrigerator~\cite{Burnett_2019}.

The main advantage of these FNN approaches is their ability to classify multi-qubit states with much higher accuracy than traditional methods, thanks to their use of nonlinear processing and ability to mitigate crosstalk. However, they require a large number of multiplications and nonlinear operations, which demand significant hardware and computational resources. This makes them challenging to implement on modern control hardware like FPGAs, limiting their scalability to larger qubit systems. Additionally, the required computations introduce additional latency, increasing the risk of decoherence.

In one implementation by Lienhard~\textit{et al.}~\cite{Lienhard_2022}, the inputs to the FNN are the raw $\mathcal{I}_n$ and $\mathcal{Q}_n$ at all times $t_n$, eliminating the need for demodulation. Despite this simplification, evaluating the model consumes most of the computational resources on a high-end FPGA~\cite{Satvik_2023}. Moreover, one-hot encoding causes the output size to grow exponentially with the number of qubits, making it impractical for larger qubit systems.

Satvik~\textit{et al.}~\cite{Satvik_2023} addressed some of these challenges by applying a matched filter to the data, reducing the number of inputs, before feeding the signal to a compact neural network. While this lowers the complexity, it reduces the classification fidelity. To improve the qubit-state-discrimination fidelity, they introduced a relaxation-matched filter (RMF), which detects qubit relaxation from the excited to the ground state. This requires manually identifying relaxation events to engineer a training dataset.

Gautam~\textit{et al.}~\cite{Gautam_2024} simplified FNNs by reducing model size and applying quantization, resulting in an evaluation time under $50\,\text{ns}$ on an FPGA. Still, this constitutes a non-negligible fraction of the total readout time, increasing the risk of decoherence.

Mude~\textit{et al.}~\cite{Mude_2025} expanded on Ref.~\cite{Satvik_2023} by introducing an excitation-matched filter (EMF) followed by a compact neural network to improve fidelity on a three-state classification task. However, identifying excitation events and training the model are computationally intensive, making real-time training impractical.

Several of the works that implement FNNs~\cite{Lienhard_2022,Satvik_2023,Gautam_2024} have also explored the use of linear support vector machines (SVMs), a type of ML algorithm that identifies an optimal hyperplane to separate the classes in the data. The hyperplane is defined by the points closest to it, known as support vectors. Finding these is costly for large datasets, preventing real-time updates. The primary advantage of SVMs is their short evaluation time, with Gautam~\textit{et al.}~\cite{Gautam_2024} reporting a latency of under $6\,\text{ns}$ for their FPGA implementation. 

\subsection{ \label{sec: Single Qubit Experimental Setup} Single-Qubit Dataset}

We apply the NG-RC algorithm to three single-qubit datasets analyzed by Khan~\textit{et al.}~\cite{Khan_2024}. Each dataset consists of measurements of a single qubit (labeled either A or B) dispersively coupled to a stripline resonator. 

The first dataset uses qubit A, which has a relatively short lifetime $T_{1} = 9.5\,\mu\text{s}$, and a high $\chi / \kappa \approx 0.78$, where $\kappa$ is the cavity loss rate, increasing its sensitivity to dephasing~\cite{Schuster_2005,Khan_2024}. These factors make accurate qubit-state-discrimination difficult. The rectangular readout pulse duration and amplitude are varied for each set of $M$ measurements.

The second dataset uses qubit B, which has the highest $T_{1}$ of $73 \,\mu\text{s}$ and a lower $\chi / \kappa \approx 0.20$. This data was collected with a fixed moderate rectangular readout pulse amplitude, and the phase between the Josephson parametric amplifier (JPA) pump and readout pulse was varied after every set of $M$ measurements.

The last dataset also uses qubit B but under different measurement conditions, which shortened $T_{1}$ to $60 \,\mu\text{s}$. Here, stronger readout pulses induce qubit-state transitions to higher energies, creating a challenging three-state classification problem for which an optimal linear filter is not known~\cite{Khan_2024}.

\subsection{ \label{sec: Multi-Qubit Experimental Setup} Multi-Qubit Dataset}

We also use the NG-RC algorithm to discriminate the states of a five-qubit processor used by Lienhard~\textit{et al.}~\cite{Lienhard_2022}. Here, the lifetimes range from $7 \,\mu\text{s}$ to $40 \,\mu\text{s}$. Each qubit is dispersively coupled to its own coplanar waveguide resonator, and the resonators are inductively coupled to a shared Purcell-filtered bus, enabling frequency-multiplexed readout.


This dataset was collected by sequentially initializing the five qubits in each of the 32 computational-basis states and performing 50,000 readout measurements per configuration, each lasting $2 \,\mu\text{s}$. Here, only the first $1 \,\mu\text{s}$ of each measurement is used to match Refs.~\cite{Lienhard_2022,Satvik_2023}, with 15,000 measurements per configuration reserved for training the algorithms and 35,000 for evaluating their performance.


\section{\label{sec:Next-Generation Reservoir Computer Design} Next-Generation Reservoir Computer Design}

The NG-RC, shown previously in Fig.~\ref{fig:storyboard}, is an ML model well adapted for processing time-series data. The nonlinear functions can take on any form, but are often taken as monomials of the input variables at different times (see $\mathcal{O}_{nonlin,m}$ in Fig.~\ref{fig:storyboard}). This enables the model to detect correlations between variables at different times~\cite{Boyd_1985}, and maps the inputs to a higher-dimensional space to enhance its processing capacity. An advantage of this model is that the nonlinear features can be computed in parallel to reduce the latency, unlike in neural networks, which must evaluate each layer before continuing. 

One downside of using monomials is that the number of features grows exponentially with the maximum polynomial degree. One method of reducing the number of monomials is to average $\mathcal{I}_n$ and $\mathcal{Q}_n$ over non-overlapping temporal windows of size $w$ and computing the polynomial features from the averaged signals $I_{n'}$ and $Q_{n'}$. There is a trade-off between model fidelity and the size $w$ as we discuss in Sec.~\ref{sec:Results}. Another approach is to use nonlinear system identification techniques~\cite{Wei_2004,Kent_2024}, which selects the most important terms to retain. This is discussed in Appendix~\ref{sec:System Identification}.

As shown in Fig.~\ref{fig:storyboard}, the linear and nonlinear features are concatenated into the total feature vector $\mathcal{O}_{total,m}$. This vector is multiplied by the weights $W_{out}$ found in the training phase, further discussed in Appendix~\ref{sec:Single-Qubit Model: Training and Testing}, producing an output $\hat{y}_m$. For binary classification tasks, such as the first two single-qubit datasets, a threshold is applied to obtain the predicted state $\hat{y}'_m \in \{0,1\}$. For the three-state classification problem in the third dataset, a one-hot encoding is used for $\hat{\mathbf{y}}_m$, and the predicted state $\hat{y}'_m$ is obtained by applying the \textit{argmax} function, which assigns the highest probability to the largest output component.

For the multi-qubit algorithm, the readout signals are digitally demodulated to extract $(\mathcal{I}_n$, $\mathcal{Q}_n)$ for each qubit. A binary mask sets these values to zero past a specific time for each qubit, where the signal no longer provides useful information about the qubit state. Next, the number of linear features is reduced by averaging over $w$. We construct a nonlinear feature vector from $I_{n'}$ and $Q_{n'}$ across all qubits and time windows, enabling the model to account for crosstalk. To keep the feature set manageable when scaling to larger systems, correlations can be limited to qubits with nearby resonator frequencies, where crosstalk is strongest.

We train five distinct NG-RC models, each predicting a single qubit's state. Like the other binary state classifiers, the output of each model is a single value $\hat{y}_m$, which is discriminated to yield a binary state $0$ or $1$, and a discriminator threshold optimized to maximize the fidelity. This avoids the exponential growth of the output size seen in Ref.~\cite{Lienhard_2022}.

\section{\label{sec:Readout Measurement Metrics} Qubit-State Readout Metrics}

We use several metrics to compare the performance of the NG-RC algorithm with other methods. For all datasets, we use the qubit-state-discrimination fidelity $\mathcal{F}$, which is the ratio of the number of correct classifications to the total number of classifications. This metric ranges from 0 to 1, with values closer to 1 indicating higher classification accuracy. 

\begin{figure*}[t]
\includegraphics[width=0.95\textwidth]{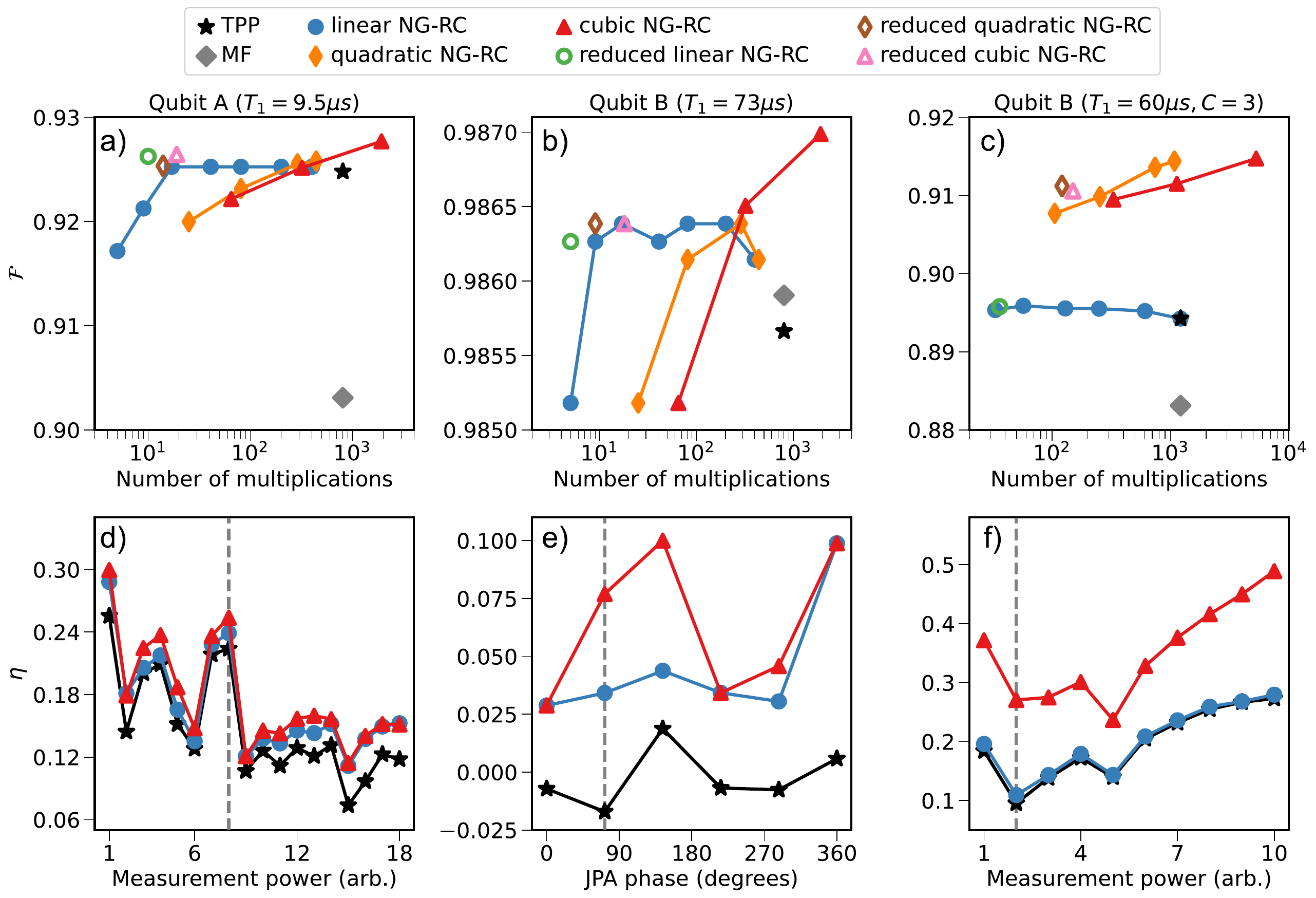}
\caption{\label{fig:single_qubit_fidelity_and_infidelity_reduction} 
Classification fidelity and infidelity reduction for single-qubit datasets. The matched filter (MF) and temporal postprocessor (TPP) results are taken from Ref.~\cite{Khan_2024}; all other data points are from this work. The points labeled ``reduced'' use system identification. (a)-(c) Classification fidelity as a function of the number of features in the model for each of the three datasets, where the measurement conditions correspond to the dashed gray lines in (d)-(f). Shown is the infidelity reduction compared to a MF as a function of the measurement conditions. In the bottom row, the black stars represent the TPP, the blue circles denote the best linear NG-RC, and the red triangles indicate the best nonlinear NG-RC.}
\end{figure*}

Another useful metric is the infidelity reduction, which is defined as the relative decrease in classification error for a ML method compared to the matched filter, the baseline method, given by 
\begin{equation}
\label{eq:infidelity_reduction}
    \eta = \frac{(1-\mathcal{F}_{MF}) - (1 - \mathcal{F}_{ML})}{1-\mathcal{F}_{MF}}.
\end{equation}
Here, $\mathcal{F}_{MF}$ and $\mathcal{F}_{ML}$ are the fidelities of the matched filter and the ML method, respectively. Positive (negative) values of $\eta$ indicate improvement (worse performance) over the matched filter.


For multi-qubit classifications of $N_q$ qubits, we use the geometric mean classification fidelity
\begin{equation}
\label{eq:geometric mean fidelity}
    \mathcal{F}_{GM} = (\Pi_{r=1}^{N_q}\mathcal{F}_r)^{1/N_q},
\end{equation}
where $\mathcal{F}_r$ is the individual qubit fidelity. As before, values closer to 1 indicate a higher classification accuracy.

To quantify the extent to which the model accounts for crosstalk, we define the cross-fidelity  
\begin{equation}
    \mathcal{F}_{jk}^{CF} = \langle 1 - \big[ P(1_j | 0_k) + P(0_j | 1_k)\big]\rangle,
\end{equation}
where $P(1_j | 0_k)$ is the conditional probability that the model predicts qubit $j$ to be in the excited state given that qubit $k$ was prepared in the ground state, and $P(0_j | 1_k)$ is the conditional probability that the model predicts qubit $j$ to be in the ground state, given that qubit $k$ was prepared in the excited state. Cross-fidelity values closer to zero indicate more effective crosstalk mitigation.

\section{\label{sec:Results} Results}

\subsection{\label{sec:Single Qubit Results} Single-Qubit Results}

The single-qubit fidelity as a function of the number of multiplications in model evaluation is shown in Fig.~\ref{fig:single_qubit_fidelity_and_infidelity_reduction}a-c. Each panel shows fidelity curves corresponding to the measurement powers (a,c) or JPA phase (b) corresponding to the highest classification fidelity achieved across all methods. The goal is to maximize fidelity while minimizing the number of multiplications, which reflects the model's complexity. The curves with multiple points correspond to the different window sizes $w$ tested (listed in Appendix~\ref{sec:Feature Creation}), with larger $w$ corresponding to fewer multiplications. The linear NG-RC point furthest to the right corresponds to no averaging.

In (a-b), the linear NG-RC without averaging requires fewer multiplications than the TPP due to its single-output encoding, whereas TPP uses one-hot encoding. For dataset (c), both methods use one-hot encoding, resulting in equal complexity and fidelity. 

Notably, as the window size increases for the linear NG-RC (a–c), the fidelity peaks before eventually declining. Moreover, applying system identification yields the highest fidelity among the linear NG-RC models in (a) and (c), while requiring only few multiplications. As a result, the linear NG-RC achieves higher fidelity than both the TPP and MF, despite being significantly less complex.

\begin{figure*}[t]
\includegraphics[width=0.95\textwidth]{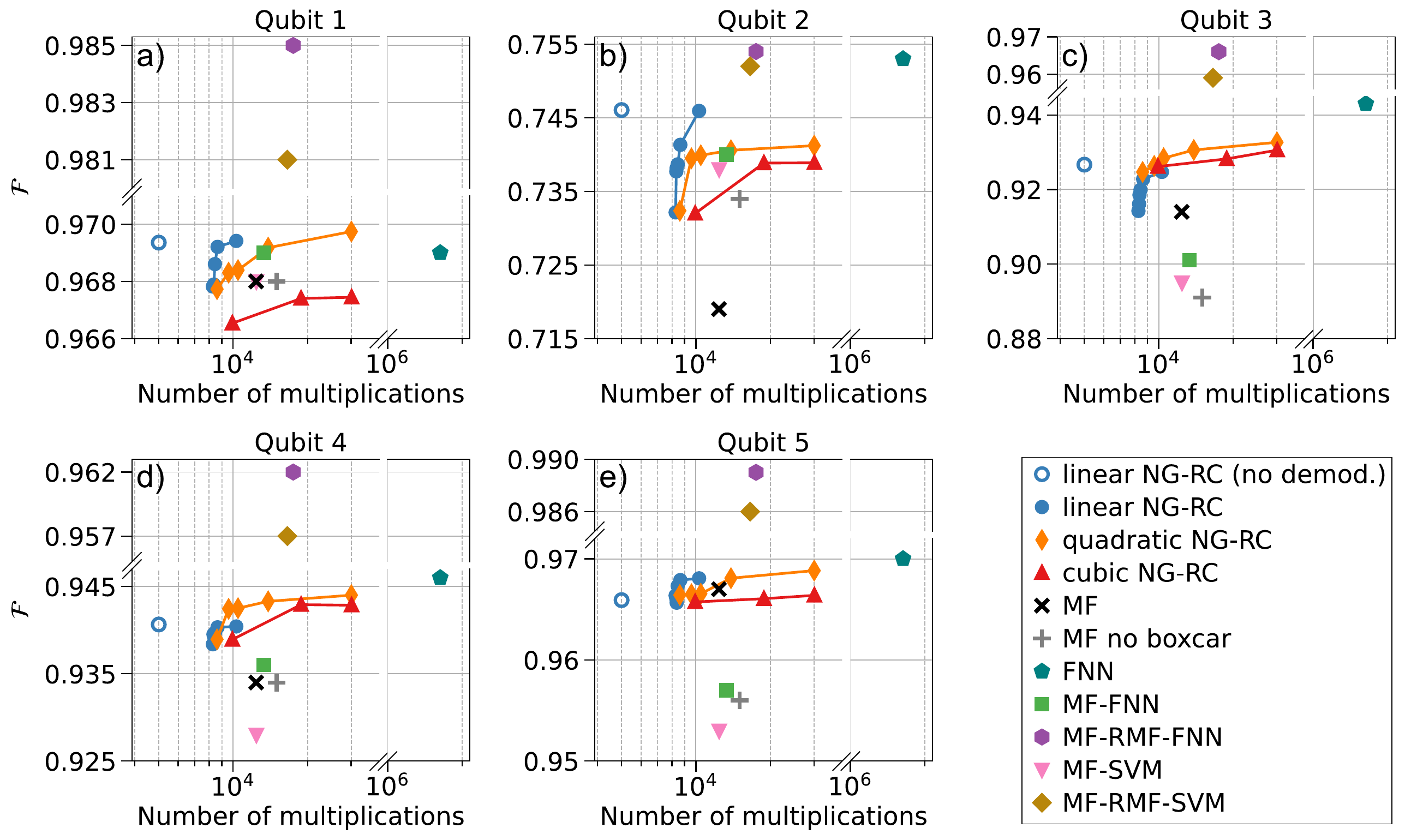}
\caption{\label{fig:multiqubit_fidelity} Classification fidelity for a 5-qubit chip. Panels (a)-(e) show the classification fidelity for qubits 1 through 5, respectively. The legend, shown to the right of panel (e), summarizes the methods compared. The MF and FNN results are taken from Ref.~\cite{Lienhard_2022}, and MF (no boxcar), MF-FNN, MF-RMF-FNN, MF-SVM, and MF-RMF-SVM are taken from Ref.~\cite{Satvik_2023}. All other results are from this work.
}
\end{figure*}

We see in (a-c) that the quadratic and cubic NG-RCs achieve a greater maximum fidelity than all other models tested. While this increase is less pronounced for (a) and (b), the nonlinear features provide a significant jump in fidelity for the three-state classification task in (c). Crucially, this fidelity gain appears even at a number of multiplications comparable to the linear models. In (a) and (b), the reduced quadratic and cubic models achieve a fidelity similar to the best linear NG-RC. In (c), they retain similar fidelity to the full nonlinear models with fewer multiplications. These results indicate that nonlinear features provide a significant increase in fidelity over linear methods for the most challenging single-qubit classification task without compromising computational efficiency.

Fig.~\ref{fig:single_qubit_fidelity_and_infidelity_reduction}d-f shows the infidelity reduction (Eq.~\ref{eq:infidelity_reduction}) relative to the MF for the TPP, the best linear NG-RC, and the best nonlinear NG-RC for all measurement powers (d,f) and JPA phases (e). The vertical dashed lines indicate the measurements that correspond to the example fidelity curves shown in Fig.~\ref{fig:single_qubit_fidelity_and_infidelity_reduction}a-c. Here, a larger infidelity reduction indicates better performance. 

In Fig.~\ref{fig:single_qubit_fidelity_and_infidelity_reduction}d-f, we see that the linear NG-RC has a higher infidelity reduction than the TPP for all measurement conditions, with the TPP performing worse than the MF for some JPA phases. The best nonlinear NG-RC always matches or outperforms the best linear NG-RC, with the highest infidelity reduction of nearly 50\% occurring for the three-state task in (f). These results highlight that the NG-RC provides fidelity gains compared to the TPP and MF across a wide range of measurement conditions.

\begin{table*}[t]
\centering
\renewcommand{\arraystretch}{1.2}
\begin{tabular}{|l|c|c|c|c|c|c|}
\hline
\textbf{Design} & \textbf{Qubit 1} & \textbf{Qubit 2} & \textbf{Qubit 3} & \textbf{Qubit 4} & \textbf{Qubit 5} & \textbf{F\textsubscript{5Q}} \\ \hline
MF~\cite{Satvik_2023} & 0.968 & 0.734 & 0.891 & 0.934 & 0.956 & 0.892 \\ \hline
MF + boxcar~\cite{Lienhard_2022} & 0.968 & 0.719 & 0.914 & 0.934 & 0.967 & 0.895 \\ \hline
MF-SVM~\cite{Satvik_2023} & 0.968 & 0.738 & 0.895 & 0.928 & 0.953 & 0.892 \\ \hline
MF-NN~\cite{Satvik_2023} & 0.969 & 0.740 & 0.901 & 0.936 & 0.957 & 0.896 \\ \hline
\textbf{Cubic NG-RC \(w=200\)} & \textbf{0.967} & \textbf{0.739} & \textbf{0.931} & \textbf{0.943} & \textbf{0.966} & \textbf{0.905} \\ \hline
\textbf{Linear NG-RC} (no demod.) & \textbf{0.969} & \textbf{0.746} & \textbf{0.927} & \textbf{0.941} & \textbf{0.966} & \textbf{0.906} \\ \hline
\textbf{Linear NG-RC \(w=10\)} & \textbf{0.969} & \textbf{0.746} & \textbf{0.925} & \textbf{0.940} & \textbf{0.968} & \textbf{0.906} \\ \hline
\textbf{Quadratic NG-RC \(w=50\)} & \textbf{0.970} & \textbf{0.741} & \textbf{0.933} & \textbf{0.944} & \textbf{0.969} & \textbf{0.907} \\ \hline
NN~\cite{Lienhard_2022} & 0.969 & 0.753 & 0.943 & 0.946 & 0.970 & 0.912 \\ \hline
MF-RMF-SVM~\cite{Satvik_2023} & 0.981 & 0.752 & 0.959 & 0.957 & 0.986 & 0.923 \\ \hline
MF-RMF-NN~\cite{Satvik_2023} & 0.985 & 0.754 & 0.966 & 0.962 & 0.989 & 0.927 \\ \hline
\end{tabular}
\caption{Individual qubit-state-discrimination fidelity and geometric mean fidelity (F\textsubscript{5Q}) for different designs.}
\label{table:qubit_fidelities}
\end{table*}

\begin{table*}[t]
\centering
\begin{tabular}{|c|c|c|c|c|c|}
\hline
\textbf{Design} & $\langle |F^{CF}_{k=j\pm1}| \rangle$ & $\langle |F^{CF}_{k=j\pm2}| \rangle$ & $\langle |F^{CF}_{k=j\pm3}| \rangle$ & $\langle |F^{CF}_{k=j\pm4}| \rangle$ & $\langle |F^{\text{CF}}| \rangle_{\Delta k}$ \\ \hline
NN~\cite{Lienhard_2022} & 0.002 & 0.005 & 0.002 & 0.0003 & 0.0023 \\ \hline
MF-RMF-NN~\cite{Satvik_2023} & 0.0031 & 0.0062 & 0.0008 & 0.0005 & 0.0027 \\ \hline
\textbf{Quadratic NG-RC} $w=50$ & \textbf{0.0037} & \textbf{0.0049} & \textbf{0.0020} & \textbf{0.0009} & \textbf{0.0029} \\ \hline
\textbf{Linear NG-RC} (no demod.) & \textbf{0.0036} & \textbf{0.0066} & \textbf{0.0024} & \textbf{0.0006} & \textbf{0.0033} \\ \hline
\textbf{Linear NG-RC} $w=10$ & \textbf{0.0048} & \textbf{0.0095} & \textbf{0.0013} & \textbf{0.0005} & \textbf{0.0040} \\ \hline
MF~\cite{Satvik_2023} & 0.0108 & 0.015 & 0.0021 & 0.0008 & 0.0072 \\ \hline
MF-NN~\cite{Satvik_2023} & 0.0071 & 0.011 & 0.003 & 0.0003 & 0.0054 \\ \hline
MF-RMF-SVM~\cite{Satvik_2023} & 0.011 & 0.0077 & 0.0024 & 0.0006 & 0.0054 \\ \hline
\end{tabular}
\caption{Mean absolute values of the cross-fidelity for different Hamming distances, with $\langle |F^{\text{CF}}| \rangle_{\Delta k}$ indicating the average over all qubit index separations $\Delta k = |k - k'|$.}
\label{tab:cross_fidelity}
\end{table*}

\begin{table*}[t]
\centering
\renewcommand{\arraystretch}{1.2}
\begin{tabular}{|l|c|c|c|c|}
\hline
\textbf{Design} & \textbf{\# Parameters} & \textbf{\# Multiplications} & \textbf{\# Activation Functions} & \textbf{F\textsubscript{5Q}} \\ \hline
MF~\cite{Satvik_2023} & $5 \times 10^3$ & $1.5 \times 10^4$ & 0 & 0.892 \\ \hline
MF + boxcar~\cite{Lienhard_2022} & $4.13 \times 10^3$ & $1.24 \times 10^4$ & 0 & 0.892 \\ \hline
MF-SVM~\cite{Satvik_2023} & 4160 & $1.24 \times 10^4$ & 0 & 0.892 \\ \hline
MF-NN~\cite{Satvik_2023} & 5082 & $1.33 \times 10^4$ & 62 & 0.896 \\ \hline
\textbf{Cubic NG-RC \(w=200\)} & $1.83 \times 10^4$ & $3.01 \times 10^4$ & 0 & 0.905 \\ \hline
\textbf{Linear NG-RC} (no demod.) & $5005$ & \textcolor{green}{$5005$} & 0 & 0.906 \\ \hline
\textbf{Linear NG-RC \(w=10\)} & \textcolor{green}{2075} & $1.03 \times 10^4$ & 0 & 0.906 \\ \hline
\textbf{Quadratic NG-RC \(w=50\)} & $1.83 \times 10^4$ & $3.01 \times 10^4$ & 0 & 0.907 \\ \hline
NN~\cite{Satvik_2023} & $1.63 \times 10^6$ & $1.63 \times 10^6$ & 1782 & 0.912 \\ \hline
MF-RMF-SVM~\cite{Satvik_2023} & 8315 & $1.66 \times 10^4$ & 0 & 0.923 \\ \hline
MF-RMF-NN~\cite{Satvik_2023} & 9262 & $1.75 \times 10^4$ & 62 & 0.927 \\ \hline
\end{tabular}
\caption{Computational complexity of discriminators.}
\label{table:complexity_comparison}
\end{table*}

\subsection{\label{sec:Multi Qubit Results} Multi-Qubit Results}

The fidelities for all five qubits, as a function of the number of multiplications, are shown in Fig.~\ref{fig:multiqubit_fidelity}a-e. Like the single qubit results, a curve with multiple points indicates that multiple window sizes are evaluated, with larger $w$ corresponding to fewer multiplications. The number of multiplications in each plot reflects the total required to predict all five qubit states. The goal is to maximize the fidelity while keeping the number of multiplications low.

We see that the linear NG-RC, without digital demodulation or window averaging, requires the fewest multiplications. When demodulation and averaging are included, increasing $w$ strictly reduces fidelity, though the overall performance remains comparable. Both linear NG-RC variants achieve a higher fidelity than the MF, an FNN, and an SVM, while requiring fewer multiplications.

The quadratic NG-RC achieves higher maximum fidelity than the linear NG-RC across all qubits, except for qubit 2, which has significantly lower fidelity across all methods. Notably, for qubits 3 and 4, it outperforms the linear NG-RC with a similar number of multiplications. The cubic model, which uses larger window sizes to reduce the number of multiplications, generally shows lower fidelity as a result.

Importantly, the NG-RC outperforms the baseline FNN for qubit 1, with only slightly lower fidelity for the other qubits, while requiring orders-of-magnitude fewer operations. While the MF-RMF-SVM and MF-RMF-FNN algorithms achieve higher fidelity than the baseline FNN with reduced complexity, their performance relies on the RMF~\cite{Satvik_2023}, which requires identifying relaxation events during dataset construction.

These results indicate that the NG-RC can outperform the traditional and some ML approaches while requiring fewer multiplications. In addition, we conclude that including nonlinear terms in the model can result in an increase in qubit-state-discrimination fidelity without substantial increases in complexity.

The maximum fidelity for each qubit and the geometric mean fidelity (Eq.~\ref{eq:geometric mean fidelity}) are summarized in Tab.~\ref{table:qubit_fidelities}, in agreement with the results discussed above.

The cross-fidelity is shown in Tab.~\ref{tab:cross_fidelity} for different Hamming distances, defined as the separation between qubit indices, with values closer to 0 indicating better crosstalk mitigation. We see that the NG-RCs have better crosstalk mitigation on average than the MF and two ML methods, with the quadratic NG-RC achieving similar performance to the neural networks in Refs.~\cite{Lienhard_2022,Satvik_2023}. We attribute this to cross-terms between individual qubits, which allow the model to learn correlations between qubits.

The computational complexity of evaluating the different models is summarized in Tab.~\ref{table:complexity_comparison}. The goal is to minimize the number of trainable parameters, multiplications, and nonlinear activation functions while maximizing the geometric mean fidelity. For models that have an MF or RMF as input, the parameters needed to learn the MF are counted as well.

The model that requires the fewest number of parameters is the linear NG-RC with an averaging window size of 10 time steps. This is orders of magnitude fewer than the other approaches, despite having a higher accuracy than traditional methods and those with MF pre-processing. We also see that the number of parameters for the quadratic and cubic NG-RC with the highest fidelity is large, but still fewer than the baseline NN~\cite{Lienhard_2022}.

The baseline NN~\cite{Lienhard_2022} requires a large number of nonlinear activation functions, which greatly increases the resources required to implement the model. While the other neural network approaches require fewer nonlinear function evaluations, they are still costly to implement. In contrast, both the traditional and NG-RC approaches avoid nonlinear activation functions entirely.

While Tab.~\ref{table:complexity_comparison} does not capture training complexity, the NG-RC and the traditional algorithms are significantly simpler to train than an FNN or SVM because they do not require nonlinear optimization. In Appendix~\ref{sec:Single-Qubit Model: Training and Testing}, we discuss how the NG-RC training can be performed directly on an FPGA in near real time.

\section{\label{sec:Discussion} Discussion}

We show that an NG-RC can perform qubit-state classification with higher fidelity than traditional methods and other ML approaches while requiring fewer operations. Additionally, we show that nonlinear terms can be included to increase fidelity further without significantly increasing the complexity of the model. Lastly, we show that system identification techniques can be used to further reduce model complexity.

Based on previous work implementing an SVM~\cite{Gautam_2024}, which has a similar evaluation complexity to the NG-RC, we expect that our approach can be evaluated on control hardware such as an FPGA in a few nanoseconds. Furthermore, the batch-training process for our approach is simple enough to be implemented in near real time to account for drifts in the system. Future work will add information about relaxation events to the feature vector, which we expect will increase the fidelity even further.

\begin{acknowledgments}
This work is supported by the Air Force Office of Scientific Research (AFOSR) Award No. FA9550-22-1-0203. B.L. is supported by Postdoc.Mobility Fellowship grant \#P500PT\_211060. We gratefully acknowledge discussions of this work with Saeed A. Khan, Boris Mesits, Hakan E. Türeci, and Michael Hatridge.
\end{acknowledgments}

\appendix

\section{\label{sec:Methods} Methods}

\subsection{\label{sec:Feature Creation} Feature Creation}

For the single-qubit datasets corresponding to Fig.~\ref{fig:single_qubit_fidelity_and_infidelity_reduction}a–c, the number of time steps (or $\mathcal{I}\mathcal{Q}$ pairs) per readout is 200, 200, and 202, respectively.

To reduce the number of features, data is averaged within non-overlapping time windows of fixed sizes $w$ = 1, 2, 5, 10, 25, 50, and 100 for the linear NG-RC; 20, 25, 50, and 100 for the quadratic NG-RC; 25, 50, and 100 for the cubic NG-RC; 5, 10, 20, 25, 50, and 100 for the linear NG-RC with system identification; 20, 25, 50, and 100 for the quadratic NG-RC with system identification; and 25, 50, and 100 for the cubic NG-RC with system identification. For the system identification results, only the window size that maximizes the fidelity is shown for the linear, quadratic, and cubic models in Fig.~\ref{fig:single_qubit_fidelity_and_infidelity_reduction}a-c, for readability.

For each dataset and each set of measurement conditions, 50\% of the data is used for training the models, and 50\% for testing the models. The quadratic NG-RCs contain the linear features, the quadratic monomials, and a constant. The cubic NG-RCs augments the quadratic model by including the cubic monomials. 

For the multi-qubit datasets, the readout signal, which originally contained 500 time steps, is digitally demodulated. To improve classification performance, a boxcar filter is used to isolate the part of the signal most relevant for each qubit. As defined in Eq.~\ref{eq:boxcar}, the boxcar involves a sum over selected time steps. In our case, it acts as a mask that keeps values within a fixed time range and sets all others to zero. Starting with the optimal boxcars from Lienhard~\textit{et al.}~\cite{Lienhard_2022}, we shift each boxcar by a single time step until the testing fidelity is maximized. The resulting boxcars begin at the first time step and end at time steps 500, 500, 282, 479, and 295 for the five qubits, respectively. In contrast, no boxcar filters are applied for the linear NG-RC that use the raw, intermediate frequency signals. 

After applying the boxcars, the individual signals are averaged with $w$ = 10, 50, 100, 200, 250, and 500 for the linear NG-RC; 50, 100, 200, 250, and 500 for the quadratic NG-RC; and 200, 250, and 500 for the cubic NG-RC.  For all experiments, only a single value of $w$ is used for each model, but a combination of window sizes could be used to maximize performance. In the multi-qubit case, an individual NG-RC is trained to predict the state of a single qubit while having access to features from the other qubit signals.

Throughout this work, we avoid recomputing quadratic terms used in the construction of cubic terms, thereby reducing the number of multiplications needed to evaluate the cubic NG-RC models.

\subsection{\label{sec:System Identification} System Identification}

To perform the system identification on the single qubit datasets, we use the package \texttt{SysIdentPy} (version 0.3.3)~\cite{SysIdentPy} with the \texttt{"NFIR"} model type, \texttt{"ridge\_regression"} estimator, and the \texttt{Polynomial} basis function. We perform a feature search for multiple values of \texttt{ridge\_param}, as discussed in the next subsection. For each NG-RC type, we truncate \texttt{n\_terms} when the difference between successive information values drops below 1\% of the maximum range to ensure the model remains compact. For the linear NG-RC, we use \texttt{degree=1} and the maximum \texttt{n\_info\_values} for the information criteria search. For both quadratic (\texttt{degree=2}) and cubic (\texttt{degree=3}) NG-RC, we use \texttt{n\_info\_values=30} to balance performance and evaluation time. 

\subsection{\label{sec:Single-Qubit Model: Training and Testing} Single-Qubit Model: Training and Testing}

During the training phase, the feature vectors $\mathcal{O}_{total,m} \in \mathbb{R}^{N_f \times 1}$ from each measurement $m$ in the training dataset are concatenated to form the matrix $\mathcal{O}_{total} \in \mathbb{R}^{N_f \times M}$. $N_f$ is the length of each feature vector and $M$ is the total number of measurements. Similarly, the ground truth labels for each measurement $y_{m} \in \mathbb{R}^{d \times 1}$ in the training dataset are concatenated to form the matrix $\mathbf{Y} \in \mathbb{R}^{d \times M}$, where the dimension $d$ of the output depends on whether one-hot encoding or threshold-based outputs are used. The weights $W_{out}$ are computed using regularized linear regression with an L2 regularization term, known as ridge regression~\cite{Vogel_2002}, given by
\begin{equation}
    W_{out} = \mathbf{Y} \mathcal{O}_{total}^\top \left( \mathcal{O}_{total} \mathcal{O}_{total}^\top + \alpha \mathbf{I} \right)^{-1}
\label{eq:ridge_regression}
\end{equation}
where $\alpha$ is the regularization parameter. Larger values of $\alpha$ penalize the size of the weights, and ensures that $\mathcal{O}_{total} \mathcal{O}_{total}^\top$ is not ill-conditioned. 

For each of the single qubit datasets and measurement conditions, we compute Eq.~\ref{eq:ridge_regression} for all measurements $m$ in the training dataset, with $\alpha$ values logarithmically spaced between $10^{-7}$ to $10^{-1}$, as well as $\alpha=0$. We then evaluate the performance of each of these weights by predicting $Y = W_{out} \mathcal{O}_{total}$ for each $\alpha$, where $Y$ contains the prediction for every measurement $m$ in the testing dataset. If the model has a single output (rather than one-hot encoding), a threshold is applied to determine the predicted state. This threshold is searched within the range $[0,1]$ in steps of 0.01. The combination of $\alpha$ and threshold that achieve the highest fidelity on the testing dataset is selected. If the output is one-hot encoded, predictions are made by choosing the index of $\mathbf{y}_{m}$ that has the highest value. In that case, the $\alpha$ value is chosen that maximizes the fidelity on the testing set, without performing a threshold search.

\subsection{\label{sec:Multi Qubit Model: Training and Testing} Multi-Qubit Model: Training and Testing}

Due to the significantly larger size of the multi-qubit dataset, the matrix multiplications and inversions in Eq.~\ref{eq:ridge_regression} are too costly to evaluate using the entire training dataset at once. Therefore, we perform the training in batches by using a variation of sequential ridge regression originally described by~\cite{Hertz_1991}. 

We begin with the solution to ordinary least squares, which is equivalent to Eq.~\ref{eq:ridge_regression} for $\alpha=0$, given by
\begin{equation}
    W_{out} = \mathbf{Y} \mathcal{O}_{total}^\top \left( \mathcal{O}_{total} \mathcal{O}_{total}^\top \right)^{-1},
\label{eq:OLS}
\end{equation} 
and we express the matrix multiplications as sums by writing
\begin{equation}
    W_{out} = \sum_{l=1}^{K} \tilde{Y}_l \mathcal{\tilde{O}}_l^T \left( \sum_{l=1}^{K} \mathcal{\tilde{O}}_l \mathcal{\tilde{O}}_l^T \right)^{-1}.
\label{eq:OLS_sum}
\end{equation} 
Here, $\tilde{Y}_l \in \mathbb{R}^{d \times k}$, $k$ is number of measurements in each batch, $K$ is the total number of batches, and $\mathcal{\tilde{O}}_l \in \mathbb{R}^{k \times N_f}$ is the feature vector for batch $l$. These sums are equivalent to iterating the maps
\begin{align}
Q_{l+1} &= Q_{l} + \tilde{Y}_{l+1} \mathcal{\tilde{O}}_{l+1}^T \quad \in (d \times N_f), 
\label{eq:OLS_QP_Q} \\
P_{l+1} &= P_{l} + \mathcal{\tilde{O}}_{l+1} \mathcal{\tilde{O}}_{l+1}^T \quad \in (N_f \times N_f), 
\label{eq:OLS_QP_P} \\
W_{\text{out}, l+1} &= Q_{l+1} \left(P_{l+1}\right)^{-1}.
\label{eq:OLS_QP_W}
\end{align}
In this iterative approach, the regularization parameter is applied through the initial conditions by taking
\begin{align}
Q_0 &= 0, \\
P_0 &= \alpha \mathbf{I} \quad \in (N_f \times N_f).
\label{eq:sequential_ridge_initial_conditions}
\end{align}
We see that this approach requires the inverse of an $(N_f \times N_f)$ matrix, which can become inefficient when the number of features is large. However, this approach is increasingly efficient when the batch size $k \gg N_f$.

The original approach by Hertz~\cite{Hertz_1991} updates the weights using a single measurement ($k=1$), which eliminates the need for matrix inversion entirely by applying the Sherman-Morrison Identity~\cite{Press_2007}. This approach is likely the most efficient option for training a discriminator in real time on an FPGA. However, processing multiple measurements simultaneously is often more computationally efficient than handling them one at a time when performing the training in Python on a desktop computer.

Another alternative is to use the Woodbury matrix identity~\cite{Higham_2002} to modify the mapping to require an inverse of a $k \times k$ matrix, which is inefficient for a large number of measurements in each batch. In this work, we take $k=32,000$ and the number of features is typically less than a few thousand, making Eq.~\ref{eq:OLS_QP_W} the most efficient.

The training set contains 15,000 measurements per possible state so that there are 15 total batches used to train the model. In the first batch, we initialize Eqs.~\ref{eq:sequential_ridge_initial_conditions} for regularization parameters logarithmically spaced between $10^{-7}$ to $10^3$, as well as $\alpha=0$. We iterate through each batch, computing Eq.~\ref{eq:OLS_QP_W} for each $\alpha$, and save each copy of the weights for the testing phase.

We also perform the testing phase in batches by computing $Y = W_{out} \mathcal{O}_{total}$ for $k$ measurements in each batch. For each qubit, we select the $\alpha$ and threshold between 0 and 1 in steps of 0.01 that maximizes the classification fidelity on the full testing set.

\bibliography{main}

\begin{thebibliography}{28}%
\makeatletter
\providecommand \@ifxundefined [1]{%
 \@ifx{#1\undefined}
}%
\providecommand \@ifnum [1]{%
 \ifnum #1\expandafter \@firstoftwo
 \else \expandafter \@secondoftwo
 \fi
}%
\providecommand \@ifx [1]{%
 \ifx #1\expandafter \@firstoftwo
 \else \expandafter \@secondoftwo
 \fi
}%
\providecommand \natexlab [1]{#1}%
\providecommand \enquote  [1]{``#1''}%
\providecommand \bibnamefont  [1]{#1}%
\providecommand \bibfnamefont [1]{#1}%
\providecommand \citenamefont [1]{#1}%
\providecommand \href@noop [0]{\@secondoftwo}%
\providecommand \href [0]{\begingroup \@sanitize@url \@href}%
\providecommand \@href[1]{\@@startlink{#1}\@@href}%
\providecommand \@@href[1]{\endgroup#1\@@endlink}%
\providecommand \@sanitize@url [0]{\catcode `\\12\catcode `\$12\catcode `\&12\catcode `\#12\catcode `\^12\catcode `\_12\catcode `\%12\relax}%
\providecommand \@@startlink[1]{}%
\providecommand \@@endlink[0]{}%
\providecommand \url  [0]{\begingroup\@sanitize@url \@url }%
\providecommand \@url [1]{\endgroup\@href {#1}{\urlprefix }}%
\providecommand \urlprefix  [0]{URL }%
\providecommand \Eprint [0]{\href }%
\providecommand \doibase [0]{https://doi.org/}%
\providecommand \selectlanguage [0]{\@gobble}%
\providecommand \bibinfo  [0]{\@secondoftwo}%
\providecommand \bibfield  [0]{\@secondoftwo}%
\providecommand \translation [1]{[#1]}%
\providecommand \BibitemOpen [0]{}%
\providecommand \bibitemStop [0]{}%
\providecommand \bibitemNoStop [0]{.\EOS\space}%
\providecommand \EOS [0]{\spacefactor3000\relax}%
\providecommand \BibitemShut  [1]{\csname bibitem#1\endcsname}%
\let\auto@bib@innerbib\@empty
\bibitem [{\citenamefont {Shor}(1994)}]{Shor_1994}%
  \BibitemOpen
  \bibfield  {author} {\bibinfo {author} {\bibfnamefont {P.}~\bibnamefont {Shor}},\ }\bibfield  {title} {\bibinfo {title} {Algorithms for quantum computation: discrete logarithms and factoring},\ }in\ \href {https://doi.org/10.1109/SFCS.1994.365700} {\emph {\bibinfo {booktitle} {Proceedings 35th Annual Symposium on Foundations of Computer Science}}}\ (\bibinfo {year} {1994})\ pp.\ \bibinfo {pages} {124--134}\BibitemShut {NoStop}%
\bibitem [{\citenamefont {Bernstein}\ and\ \citenamefont {Lange}(2017)}]{Bernstein_2017}%
  \BibitemOpen
  \bibfield  {author} {\bibinfo {author} {\bibfnamefont {D.~J.}\ \bibnamefont {Bernstein}}\ and\ \bibinfo {author} {\bibfnamefont {T.}~\bibnamefont {Lange}},\ }\bibfield  {title} {\bibinfo {title} {Post-quantum cryptography},\ }\href {https://doi.org/10.1038/nature23461} {\bibfield  {journal} {\bibinfo  {journal} {Nature}\ }\textbf {\bibinfo {volume} {549}},\ \bibinfo {pages} {188} (\bibinfo {year} {2017})}\BibitemShut {NoStop}%
\bibitem [{\citenamefont {Gidney}(2025)}]{Gidney_2025}%
  \BibitemOpen
  \bibfield  {author} {\bibinfo {author} {\bibfnamefont {C.}~\bibnamefont {Gidney}},\ }\href {https://arxiv.org/abs/2505.15917} {\bibinfo {title} {How to factor 2048 bit rsa integers with less than a million noisy qubits}} (\bibinfo {year} {2025}),\ \Eprint {https://arxiv.org/abs/2505.15917} {arXiv:2505.15917} \BibitemShut {NoStop}%
\bibitem [{\citenamefont {Cao}\ \emph {et~al.}(2018)\citenamefont {Cao}, \citenamefont {Romero},\ and\ \citenamefont {Aspuru-Guzik}}]{Cao_2018}%
  \BibitemOpen
  \bibfield  {author} {\bibinfo {author} {\bibfnamefont {Y.}~\bibnamefont {Cao}}, \bibinfo {author} {\bibfnamefont {J.}~\bibnamefont {Romero}},\ and\ \bibinfo {author} {\bibfnamefont {A.}~\bibnamefont {Aspuru-Guzik}},\ }\bibfield  {title} {\bibinfo {title} {Potential of quantum computing for drug discovery},\ }\href@noop {} {\bibfield  {journal} {\bibinfo  {journal} {IBM J. Res. Dev.}\ }\textbf {\bibinfo {volume} {62}},\ \bibinfo {pages} {6} (\bibinfo {year} {2018})}\BibitemShut {NoStop}%
\bibitem [{\citenamefont {Blunt}\ \emph {et~al.}(2022)\citenamefont {Blunt}, \citenamefont {Camps}, \citenamefont {Crawford}, \citenamefont {Izs{\'a}k}, \citenamefont {Leontica}, \citenamefont {Mirani}, \citenamefont {Moylett}, \citenamefont {Scivier}, \citenamefont {Sunderhauf} \emph {et~al.}}]{Blunt_2022}%
  \BibitemOpen
  \bibfield  {author} {\bibinfo {author} {\bibfnamefont {N.~S.}\ \bibnamefont {Blunt}}, \bibinfo {author} {\bibfnamefont {J.}~\bibnamefont {Camps}}, \bibinfo {author} {\bibfnamefont {O.}~\bibnamefont {Crawford}}, \bibinfo {author} {\bibfnamefont {R.}~\bibnamefont {Izs{\'a}k}}, \bibinfo {author} {\bibfnamefont {S.}~\bibnamefont {Leontica}}, \bibinfo {author} {\bibfnamefont {A.}~\bibnamefont {Mirani}}, \bibinfo {author} {\bibfnamefont {A.~E.}\ \bibnamefont {Moylett}}, \bibinfo {author} {\bibfnamefont {S.~A.}\ \bibnamefont {Scivier}}, \bibinfo {author} {\bibfnamefont {C.}~\bibnamefont {Sunderhauf}}, \emph {et~al.},\ }\bibfield  {title} {\bibinfo {title} {Perspective on the current state-of-the-art of quantum computing for drug discovery applications},\ }\href {https://doi.org/10.1021/acs.jctc.2c00574} {\bibfield  {journal} {\bibinfo  {journal} {J. Chem. Theory Comput.}\ }\textbf {\bibinfo {volume} {18}},\ \bibinfo {pages} {7001} (\bibinfo {year} {2022})}\BibitemShut {NoStop}%
\bibitem [{\citenamefont {Farhi}\ \emph {et~al.}(2014)\citenamefont {Farhi}, \citenamefont {Goldstone},\ and\ \citenamefont {Gutmann}}]{Farhi_2014}%
  \BibitemOpen
  \bibfield  {author} {\bibinfo {author} {\bibfnamefont {E.}~\bibnamefont {Farhi}}, \bibinfo {author} {\bibfnamefont {J.}~\bibnamefont {Goldstone}},\ and\ \bibinfo {author} {\bibfnamefont {S.}~\bibnamefont {Gutmann}},\ }\href {https://arxiv.org/abs/1411.4028} {\bibinfo {title} {A quantum approximate optimization algorithm}} (\bibinfo {year} {2014}),\ \Eprint {https://arxiv.org/abs/1411.4028} {arXiv:1411.4028} \BibitemShut {NoStop}%
\bibitem [{\citenamefont {Zhou}\ \emph {et~al.}(2020)\citenamefont {Zhou}, \citenamefont {Wang}, \citenamefont {Choi}, \citenamefont {Pichler},\ and\ \citenamefont {Lukin}}]{Zhou_2020}%
  \BibitemOpen
  \bibfield  {author} {\bibinfo {author} {\bibfnamefont {L.}~\bibnamefont {Zhou}}, \bibinfo {author} {\bibfnamefont {S.-T.}\ \bibnamefont {Wang}}, \bibinfo {author} {\bibfnamefont {S.}~\bibnamefont {Choi}}, \bibinfo {author} {\bibfnamefont {H.}~\bibnamefont {Pichler}},\ and\ \bibinfo {author} {\bibfnamefont {M.~D.}\ \bibnamefont {Lukin}},\ }\bibfield  {title} {\bibinfo {title} {Quantum approximate optimization algorithm: Performance, mechanism, and implementation on near-term devices},\ }\href {https://doi.org/10.1103/PhysRevX.10.021067} {\bibfield  {journal} {\bibinfo  {journal} {Phys. Rev. X}\ }\textbf {\bibinfo {volume} {10}},\ \bibinfo {pages} {021067} (\bibinfo {year} {2020})}\BibitemShut {NoStop}%
\bibitem [{\citenamefont {Abbas}\ \emph {et~al.}(2024)\citenamefont {Abbas} \emph {et~al.}}]{optimization_nature_2024}%
  \BibitemOpen
  \bibfield  {author} {\bibinfo {author} {\bibfnamefont {A.}~\bibnamefont {Abbas}} \emph {et~al.},\ }\bibfield  {title} {\bibinfo {title} {Challenges and opportunities in quantum optimization},\ }\href {https://doi.org/10.1038/s42254-024-00770-9} {\bibfield  {journal} {\bibinfo  {journal} {Nat. Rev. Phys.}\ }\textbf {\bibinfo {volume} {6}},\ \bibinfo {pages} {718} (\bibinfo {year} {2024})}\BibitemShut {NoStop}%
\bibitem [{\citenamefont {Shor}(1995)}]{Shor_1995}%
  \BibitemOpen
  \bibfield  {author} {\bibinfo {author} {\bibfnamefont {P.~W.}\ \bibnamefont {Shor}},\ }\bibfield  {title} {\bibinfo {title} {Scheme for reducing decoherence in quantum computer memory},\ }\href@noop {} {\bibfield  {journal} {\bibinfo  {journal} {Physical review A}\ }\textbf {\bibinfo {volume} {52}},\ \bibinfo {pages} {R2493} (\bibinfo {year} {1995})}\BibitemShut {NoStop}%
\bibitem [{\citenamefont {Fowler}\ \emph {et~al.}(2012)\citenamefont {Fowler}, \citenamefont {Mariantoni}, \citenamefont {Martinis},\ and\ \citenamefont {Cleland}}]{Fowler_2012}%
  \BibitemOpen
  \bibfield  {author} {\bibinfo {author} {\bibfnamefont {A.~G.}\ \bibnamefont {Fowler}}, \bibinfo {author} {\bibfnamefont {M.}~\bibnamefont {Mariantoni}}, \bibinfo {author} {\bibfnamefont {J.~M.}\ \bibnamefont {Martinis}},\ and\ \bibinfo {author} {\bibfnamefont {A.~N.}\ \bibnamefont {Cleland}},\ }\bibfield  {title} {\bibinfo {title} {Surface codes: Towards practical large-scale quantum computation},\ }\href {https://doi.org/10.1103/PhysRevA.86.032324} {\bibfield  {journal} {\bibinfo  {journal} {Phys. Rev. A}\ }\textbf {\bibinfo {volume} {86}},\ \bibinfo {pages} {032324} (\bibinfo {year} {2012})}\BibitemShut {NoStop}%
\bibitem [{\citenamefont {{Google Quantum AI and Collaborators}}(2024)}]{Google_2024}%
  \BibitemOpen
  \bibfield  {author} {\bibinfo {author} {\bibnamefont {{Google Quantum AI and Collaborators}}},\ }\bibfield  {title} {\bibinfo {title} {Quantum error correction below the surface code threshold},\ }\bibfield  {journal} {\bibinfo  {journal} {Nature}\ }\href {https://doi.org/10.1038/s41586-024-08449-y} {10.1038/s41586-024-08449-y} (\bibinfo {year} {2024})\BibitemShut {NoStop}%
\bibitem [{\citenamefont {Khan}\ \emph {et~al.}(2024)\citenamefont {Khan}, \citenamefont {Kaufman}, \citenamefont {Mesits}, \citenamefont {Hatridge},\ and\ \citenamefont {T\"ureci}}]{Khan_2024}%
  \BibitemOpen
  \bibfield  {author} {\bibinfo {author} {\bibfnamefont {S.~A.}\ \bibnamefont {Khan}}, \bibinfo {author} {\bibfnamefont {R.}~\bibnamefont {Kaufman}}, \bibinfo {author} {\bibfnamefont {B.}~\bibnamefont {Mesits}}, \bibinfo {author} {\bibfnamefont {M.}~\bibnamefont {Hatridge}},\ and\ \bibinfo {author} {\bibfnamefont {H.~E.}\ \bibnamefont {T\"ureci}},\ }\bibfield  {title} {\bibinfo {title} {Practical trainable temporal postprocessor for multistate quantum measurement},\ }\href {https://doi.org/10.1103/PRXQuantum.5.020364} {\bibfield  {journal} {\bibinfo  {journal} {PRX Quantum}\ }\textbf {\bibinfo {volume} {5}},\ \bibinfo {pages} {020364} (\bibinfo {year} {2024})}\BibitemShut {NoStop}%
\bibitem [{\citenamefont {Lienhard}\ \emph {et~al.}(2022)\citenamefont {Lienhard}, \citenamefont {Veps\"al\"ainen}, \citenamefont {Govia}, \citenamefont {Hoffer}, \citenamefont {Qiu}, \citenamefont {Rist\`e}, \citenamefont {Ware}, \citenamefont {Kim}, \citenamefont {Winik}, \citenamefont {Melville} \emph {et~al.}}]{Lienhard_2022}%
  \BibitemOpen
  \bibfield  {author} {\bibinfo {author} {\bibfnamefont {B.}~\bibnamefont {Lienhard}}, \bibinfo {author} {\bibfnamefont {A.}~\bibnamefont {Veps\"al\"ainen}}, \bibinfo {author} {\bibfnamefont {L.~C.}\ \bibnamefont {Govia}}, \bibinfo {author} {\bibfnamefont {C.~R.}\ \bibnamefont {Hoffer}}, \bibinfo {author} {\bibfnamefont {J.~Y.}\ \bibnamefont {Qiu}}, \bibinfo {author} {\bibfnamefont {D.}~\bibnamefont {Rist\`e}}, \bibinfo {author} {\bibfnamefont {M.}~\bibnamefont {Ware}}, \bibinfo {author} {\bibfnamefont {D.}~\bibnamefont {Kim}}, \bibinfo {author} {\bibfnamefont {R.}~\bibnamefont {Winik}}, \bibinfo {author} {\bibfnamefont {A.}~\bibnamefont {Melville}}, \emph {et~al.},\ }\bibfield  {title} {\bibinfo {title} {Deep-neural-network discrimination of multiplexed superconducting-qubit states},\ }\href {https://doi.org/10.1103/PhysRevApplied.17.014024} {\bibfield  {journal} {\bibinfo  {journal} {Phys. Rev. Appl.}\ }\textbf {\bibinfo {volume} {17}},\ \bibinfo {pages} {014024} (\bibinfo {year} {2022})}\BibitemShut
  {NoStop}%
\bibitem [{\citenamefont {Maurya}\ \emph {et~al.}(2023)\citenamefont {Maurya}, \citenamefont {Mude}, \citenamefont {Oliver}, \citenamefont {Lienhard},\ and\ \citenamefont {Tannu}}]{Satvik_2023}%
  \BibitemOpen
  \bibfield  {author} {\bibinfo {author} {\bibfnamefont {S.}~\bibnamefont {Maurya}}, \bibinfo {author} {\bibfnamefont {C.~N.}\ \bibnamefont {Mude}}, \bibinfo {author} {\bibfnamefont {W.~D.}\ \bibnamefont {Oliver}}, \bibinfo {author} {\bibfnamefont {B.}~\bibnamefont {Lienhard}},\ and\ \bibinfo {author} {\bibfnamefont {S.}~\bibnamefont {Tannu}},\ }\bibfield  {title} {\bibinfo {title} {Scaling qubit readout with hardware efficient machine learning architectures},\ }in\ \href {https://doi.org/10.1145/3579371.3589042} {\emph {\bibinfo {booktitle} {Proceedings of the 50th Annual International Symposium on Computer Architecture}}},\ \bibinfo {series and number} {ISCA '23}\ (\bibinfo  {publisher} {Association for Computing Machinery},\ \bibinfo {address} {New York, NY, USA},\ \bibinfo {year} {2023})\BibitemShut {NoStop}%
\bibitem [{\citenamefont {Gautam}\ \emph {et~al.}(2024)\citenamefont {Gautam}, \citenamefont {Kalipatnapu}, \citenamefont {H}, \citenamefont {Singhal}, \citenamefont {Lienhard}, \citenamefont {Singh},\ and\ \citenamefont {Thakur}}]{Gautam_2024}%
  \BibitemOpen
  \bibfield  {author} {\bibinfo {author} {\bibfnamefont {P.~K.}\ \bibnamefont {Gautam}}, \bibinfo {author} {\bibfnamefont {S.}~\bibnamefont {Kalipatnapu}}, \bibinfo {author} {\bibfnamefont {S.}~\bibnamefont {H}}, \bibinfo {author} {\bibfnamefont {U.}~\bibnamefont {Singhal}}, \bibinfo {author} {\bibfnamefont {B.}~\bibnamefont {Lienhard}}, \bibinfo {author} {\bibfnamefont {V.}~\bibnamefont {Singh}},\ and\ \bibinfo {author} {\bibfnamefont {C.~S.}\ \bibnamefont {Thakur}},\ }\href {https://arxiv.org/abs/2407.03852} {\bibinfo {title} {Low-latency machine learning fpga accelerator for multi-qubit-state discrimination}} (\bibinfo {year} {2024}),\ \Eprint {https://arxiv.org/abs/2407.03852} {arXiv:2407.03852} \BibitemShut {NoStop}%
\bibitem [{\citenamefont {Gauthier}\ \emph {et~al.}(2021)\citenamefont {Gauthier}, \citenamefont {Bollt}, \citenamefont {Griffith},\ and\ \citenamefont {Barbosa}}]{Gauthier_2021}%
  \BibitemOpen
  \bibfield  {author} {\bibinfo {author} {\bibfnamefont {D.~J.}\ \bibnamefont {Gauthier}}, \bibinfo {author} {\bibfnamefont {E.}~\bibnamefont {Bollt}}, \bibinfo {author} {\bibfnamefont {A.}~\bibnamefont {Griffith}},\ and\ \bibinfo {author} {\bibfnamefont {W.~A.~S.}\ \bibnamefont {Barbosa}},\ }\bibfield  {title} {\bibinfo {title} {Next generation reservoir computing},\ }\href {https://doi.org/10.1038/s41467-021-25801-2} {\bibfield  {journal} {\bibinfo  {journal} {Nat. Commun.}\ }\textbf {\bibinfo {volume} {12}},\ \bibinfo {pages} {5564} (\bibinfo {year} {2021})}\BibitemShut {NoStop}%
\bibitem [{\citenamefont {Reuer}\ \emph {et~al.}(2023)\citenamefont {Reuer}, \citenamefont {Landgraf}, \citenamefont {F{\"o}sel}, \citenamefont {O'Sullivan}, \citenamefont {Beltr{\'a}n}, \citenamefont {Akin}, \citenamefont {Norris}, \citenamefont {Remm}, \citenamefont {Kerschbaum}, \citenamefont {Besse} \emph {et~al.}}]{Reuer_2023}%
  \BibitemOpen
  \bibfield  {author} {\bibinfo {author} {\bibfnamefont {K.}~\bibnamefont {Reuer}}, \bibinfo {author} {\bibfnamefont {J.}~\bibnamefont {Landgraf}}, \bibinfo {author} {\bibfnamefont {T.}~\bibnamefont {F{\"o}sel}}, \bibinfo {author} {\bibfnamefont {J.}~\bibnamefont {O'Sullivan}}, \bibinfo {author} {\bibfnamefont {L.}~\bibnamefont {Beltr{\'a}n}}, \bibinfo {author} {\bibfnamefont {A.}~\bibnamefont {Akin}}, \bibinfo {author} {\bibfnamefont {G.~J.}\ \bibnamefont {Norris}}, \bibinfo {author} {\bibfnamefont {A.}~\bibnamefont {Remm}}, \bibinfo {author} {\bibfnamefont {M.}~\bibnamefont {Kerschbaum}}, \bibinfo {author} {\bibfnamefont {J.-C.}\ \bibnamefont {Besse}}, \emph {et~al.},\ }\bibfield  {title} {\bibinfo {title} {Realizing a deep reinforcement learning agent for real-time quantum feedback},\ }\href {https://doi.org/10.1038/s41467-023-42901-3} {\bibfield  {journal} {\bibinfo  {journal} {Nat. Commun.}\ }\textbf {\bibinfo {volume} {14}},\ \bibinfo {pages} {7138} (\bibinfo {year} {2023})}\BibitemShut {NoStop}%
\bibitem [{\citenamefont {Burnett}\ \emph {et~al.}(2019)\citenamefont {Burnett}, \citenamefont {Bengtsson}, \citenamefont {Scigliuzzo}, \citenamefont {Niepce}, \citenamefont {Kudra}, \citenamefont {Delsing},\ and\ \citenamefont {Bylander}}]{Burnett_2019}%
  \BibitemOpen
  \bibfield  {author} {\bibinfo {author} {\bibfnamefont {J.~J.}\ \bibnamefont {Burnett}}, \bibinfo {author} {\bibfnamefont {A.}~\bibnamefont {Bengtsson}}, \bibinfo {author} {\bibfnamefont {M.}~\bibnamefont {Scigliuzzo}}, \bibinfo {author} {\bibfnamefont {D.}~\bibnamefont {Niepce}}, \bibinfo {author} {\bibfnamefont {M.}~\bibnamefont {Kudra}}, \bibinfo {author} {\bibfnamefont {P.}~\bibnamefont {Delsing}},\ and\ \bibinfo {author} {\bibfnamefont {J.}~\bibnamefont {Bylander}},\ }\bibfield  {title} {\bibinfo {title} {Decoherence benchmarking of superconducting qubits},\ }\href {https://doi.org/10.1038/s41534-019-0168-5} {\bibfield  {journal} {\bibinfo  {journal} {npj Quantum Inf.}\ }\textbf {\bibinfo {volume} {5}},\ \bibinfo {pages} {54} (\bibinfo {year} {2019})}\BibitemShut {NoStop}%
\bibitem [{\citenamefont {Mude}\ \emph {et~al.}(2025)\citenamefont {Mude}, \citenamefont {Maurya}, \citenamefont {Lienhard},\ and\ \citenamefont {Tannu}}]{Mude_2025}%
  \BibitemOpen
  \bibfield  {author} {\bibinfo {author} {\bibfnamefont {C.~N.}\ \bibnamefont {Mude}}, \bibinfo {author} {\bibfnamefont {S.}~\bibnamefont {Maurya}}, \bibinfo {author} {\bibfnamefont {B.}~\bibnamefont {Lienhard}},\ and\ \bibinfo {author} {\bibfnamefont {S.}~\bibnamefont {Tannu}},\ }\href {https://arxiv.org/abs/2405.08982} {\bibinfo {title} {Efficient and scalable architectures for multi-level superconducting qubit readout}} (\bibinfo {year} {2025}),\ \Eprint {https://arxiv.org/abs/2405.08982} {arXiv:2405.08982 [quant-ph]} \BibitemShut {NoStop}%
\bibitem [{\citenamefont {Schuster}\ \emph {et~al.}(2005)\citenamefont {Schuster}, \citenamefont {Wallraff}, \citenamefont {Blais}, \citenamefont {Frunzio}, \citenamefont {Huang}, \citenamefont {Majer}, \citenamefont {Girvin},\ and\ \citenamefont {Schoelkopf}}]{Schuster_2005}%
  \BibitemOpen
  \bibfield  {author} {\bibinfo {author} {\bibfnamefont {D.~I.}\ \bibnamefont {Schuster}}, \bibinfo {author} {\bibfnamefont {A.}~\bibnamefont {Wallraff}}, \bibinfo {author} {\bibfnamefont {A.}~\bibnamefont {Blais}}, \bibinfo {author} {\bibfnamefont {L.}~\bibnamefont {Frunzio}}, \bibinfo {author} {\bibfnamefont {R.-S.}\ \bibnamefont {Huang}}, \bibinfo {author} {\bibfnamefont {J.}~\bibnamefont {Majer}}, \bibinfo {author} {\bibfnamefont {S.~M.}\ \bibnamefont {Girvin}},\ and\ \bibinfo {author} {\bibfnamefont {R.~J.}\ \bibnamefont {Schoelkopf}},\ }\bibfield  {title} {\bibinfo {title} {ac stark shift and dephasing of a superconducting qubit strongly coupled to a cavity field},\ }\href {https://doi.org/10.1103/PhysRevLett.94.123602} {\bibfield  {journal} {\bibinfo  {journal} {Phys. Rev. Lett.}\ }\textbf {\bibinfo {volume} {94}},\ \bibinfo {pages} {123602} (\bibinfo {year} {2005})}\BibitemShut {NoStop}%
\bibitem [{\citenamefont {Boyd}\ and\ \citenamefont {Chua}(1985)}]{Boyd_1985}%
  \BibitemOpen
  \bibfield  {author} {\bibinfo {author} {\bibfnamefont {S.}~\bibnamefont {Boyd}}\ and\ \bibinfo {author} {\bibfnamefont {L.}~\bibnamefont {Chua}},\ }\bibfield  {title} {\bibinfo {title} {Fading memory and the problem of approximating nonlinear operators with volterra series},\ }\href {https://doi.org/10.1109/TCS.1985.1085649} {\bibfield  {journal} {\bibinfo  {journal} {IEEE Trans. Circuits Syst.}\ }\textbf {\bibinfo {volume} {32}},\ \bibinfo {pages} {1150} (\bibinfo {year} {1985})}\BibitemShut {NoStop}%
\bibitem [{\citenamefont {Wei}\ \emph {et~al.}(2004)\citenamefont {Wei}, \citenamefont {Billings},\ and\ \citenamefont {Liu}}]{Wei_2004}%
  \BibitemOpen
  \bibfield  {author} {\bibinfo {author} {\bibfnamefont {H.~L.}\ \bibnamefont {Wei}}, \bibinfo {author} {\bibfnamefont {S.~A.}\ \bibnamefont {Billings}},\ and\ \bibinfo {author} {\bibfnamefont {J.}~\bibnamefont {Liu}},\ }\bibfield  {title} {\bibinfo {title} {Term and variable selection for non-linear system identification},\ }\href {https://doi.org/10.1080/00207170310001639640} {\bibfield  {journal} {\bibinfo  {journal} {Int. J. Inj. Control}\ }\textbf {\bibinfo {volume} {77}},\ \bibinfo {pages} {86} (\bibinfo {year} {2004})}\BibitemShut {NoStop}%
\bibitem [{\citenamefont {Kent}\ \emph {et~al.}(2024)\citenamefont {Kent}, \citenamefont {Barbosa},\ and\ \citenamefont {Gauthier}}]{Kent_2024}%
  \BibitemOpen
  \bibfield  {author} {\bibinfo {author} {\bibfnamefont {R.~M.}\ \bibnamefont {Kent}}, \bibinfo {author} {\bibfnamefont {W.~A.~S.}\ \bibnamefont {Barbosa}},\ and\ \bibinfo {author} {\bibfnamefont {D.~J.}\ \bibnamefont {Gauthier}},\ }\bibfield  {title} {\bibinfo {title} {Controlling chaos using edge computing hardware},\ }\href {https://doi.org/10.1038/s41467-024-48133-3} {\bibfield  {journal} {\bibinfo  {journal} {Nat. Commun.}\ }\textbf {\bibinfo {volume} {15}},\ \bibinfo {pages} {3886} (\bibinfo {year} {2024})}\BibitemShut {NoStop}%
\bibitem [{\citenamefont {Lacerda}\ \emph {et~al.}(2020)\citenamefont {Lacerda}, \citenamefont {Andrade}, \citenamefont {Oliveira},\ and\ \citenamefont {Martins}}]{SysIdentPy}%
  \BibitemOpen
  \bibfield  {author} {\bibinfo {author} {\bibfnamefont {W.~R.}\ \bibnamefont {Lacerda}}, \bibinfo {author} {\bibfnamefont {L.~P. C.~d.}\ \bibnamefont {Andrade}}, \bibinfo {author} {\bibfnamefont {S.~C.~P.}\ \bibnamefont {Oliveira}},\ and\ \bibinfo {author} {\bibfnamefont {S.~A.~M.}\ \bibnamefont {Martins}},\ }\bibfield  {title} {\bibinfo {title} {Sysidentpy: A python package for system identification using narmax models},\ }\href {https://doi.org/10.21105/joss.02384} {\bibfield  {journal} {\bibinfo  {journal} {Journal of Open Source Software}\ }\textbf {\bibinfo {volume} {5}},\ \bibinfo {pages} {2384} (\bibinfo {year} {2020})}\BibitemShut {NoStop}%
\bibitem [{\citenamefont {Vogel}(2002)}]{Vogel_2002}%
  \BibitemOpen
  \bibfield  {author} {\bibinfo {author} {\bibfnamefont {C.~R.}\ \bibnamefont {Vogel}},\ }\href {https://doi.org/10.1137/1.9780898717570} {\emph {\bibinfo {title} {Computational methods for inverse problems}}}\ (\bibinfo  {publisher} {SIAM},\ \bibinfo {year} {2002})\BibitemShut {NoStop}%
\bibitem [{\citenamefont {Hertz}(1991)}]{Hertz_1991}%
  \BibitemOpen
  \bibfield  {author} {\bibinfo {author} {\bibfnamefont {D.}~\bibnamefont {Hertz}},\ }\bibfield  {title} {\bibinfo {title} {Sequential ridge regression},\ }\href {https://doi.org/10.1109/7.81440} {\bibfield  {journal} {\bibinfo  {journal} {IEEE Trans. Aerosp. Electron. Syst.}\ }\textbf {\bibinfo {volume} {27}},\ \bibinfo {pages} {571} (\bibinfo {year} {1991})}\BibitemShut {NoStop}%
\bibitem [{\citenamefont {Press}(2007)}]{Press_2007}%
  \BibitemOpen
  \bibfield  {author} {\bibinfo {author} {\bibfnamefont {W.~H.}\ \bibnamefont {Press}},\ }\href@noop {} {\emph {\bibinfo {title} {Numerical recipes 3rd edition: The art of scientific computing}}}\ (\bibinfo  {publisher} {Cambridge university press},\ \bibinfo {year} {2007})\BibitemShut {NoStop}%
\bibitem [{\citenamefont {Higham}(2002)}]{Higham_2002}%
  \BibitemOpen
  \bibfield  {author} {\bibinfo {author} {\bibfnamefont {N.~J.}\ \bibnamefont {Higham}},\ }\href@noop {} {\emph {\bibinfo {title} {Accuracy and stability of numerical algorithms}}}\ (\bibinfo  {publisher} {SIAM},\ \bibinfo {year} {2002})\BibitemShut {NoStop}%
\end{thebibliography}%

\end{document}